\documentclass[11pt]{article}
\usepackage[latin1]{inputenc}
\usepackage[english]{babel}
\usepackage{graphicx}
\usepackage[namelimits]{amsmath}
\usepackage{amssymb}
\usepackage{amsmath}
\usepackage{amsthm}

\begin{document}
\title{Interior Kerr Solutions With The Newman-Janis Algorithm Starting With
Static Physically Reasonable Space-Times.}
\author{Stefano Viaggiu
\\
Dipartimento di Matematica, Universit\'a di Roma ``Tor Vergata'',\\
Via della Ricerca Scientifica, 1, I-00133 Roma, Italy\\
E-mail: viaggiu@mat.uniroma2.it\\
(or: stefano.viaggiu@ax0rm1.roma1.infn.it)}
\date{\today}\maketitle
\begin{abstract}
We present a simple approach for obtaining 
Kerr interior solutions with the help
of the Newman-Janis algorithm (NJA) starting with static spacetimes describing
physically sensible interior Schwarzschild solutions. In this context, the
Darmois-Israel (DI) junction conditions are analyzed. Starting from
the incompressible Schwarzschild solution, a class of Kerr interior
solutions is presented, together with a discussion of the slowly rotating 
limit. The energy conditions are discussed for the solutions so obtained.
Finally, the NJA algorithm is applied to the static, 
anisotropic, conformally flat solutions
found by Stewart leading to interior Kerr solutions with oblate 
spheroidal boundary surfaces.
\end{abstract}

\section*{Introduction}
 
The Kerr metric \cite{1} has been one of the most important astrophysics 
discovery of the last 40 years. In paricular, this solution gives the 
exact description of all black holes born from a collapsing star.
Further, Kerr claimed that the metric describes the exterior gravitational
field of a rotating body. However, since the Kerr solution has been
discovered, no physically reasonable sources have been found.
The problem of finding possible Kerr sources is that,
in order to obtain a physically sensible mass
distribution, many restrictions must be imposed. 
First of all, the metric must be joined 
smoothly to the Kerr one on a reasonable surface
for a rotating body and the 
hydrostatics pressure
must be zero on such a surface. Moreover, the energy conditions
must
hold. Finally, the star must be a non-radiating source and in the static
limit a reasonable Schwarzschild interior metric must be obtained.
Many techniques exist in the literature to build a trial interior solution
\cite{2,3,4,5,6,7,8,9,10}. First of all, the quoted Geroch
conjecture that Kerr metric might have no sources other than a black hole
must be noticed.
However, a proof is still lacking.\\ 
In this paper we present a class of
interior solutions  that match with the Kerr one on an oblate spheroid
with the help of NJA \cite{7}. Recently \cite{11,12}, it has been proven
that anisotropic fluid interior solutions can be obtained with the NJA,
after evaluating the DI \cite{9} junction conditions on an 
oblate spheroid. Although  at first a perfect fluid
source seemed the most natural choice, an anisotropic source seems now 
to be the most appropriate. For example Florides \cite{2,3}, has been
able to match an anisotropic rotating fluid 
on an oblate spheroid with the Kerr metric up
to a fifth order of a perturbative parameter. Thus, although a perfect
fluid source cannot be excluded a priori, there seem to be no convincing 
reasons to believe that one must exist.\\
In \cite{11} the authors present a trial solution requiring that in the 
static limit, a physically reasonable Schwarzschild interior
metric is obtained. To this purpose, the Volkov Oppenheimer \cite{13} equation
must be solved because, in order to apply the NJA, we need an
explicit seed static metric. However, generally, only numerical methods
are available to solve such an equation.\\
In this paper we give a simple constructive approach  
to obtain a class of interior solutions  that could act as 
physically reasonable Kerr sources.\\ 
In section 1 we give a brief introduction to the NJA . 
In section 2 we present 
our simple constructive approach. In particular, we study the DI
junction conditions in relation to the chosen seed metric.
In section 3 we present a class of interior anisotropic solutions starting
from the incompressible perfect fluid Schwarzshild metric. Section
4 is devoted to the study of the slowly rotating limit of the solution so 
obtained. In section 5 the energy conditions and the regularity are
discussed. Finally, in section 6, starting with a static 
anisotropic conformally flat
solution found in \cite{st}, a
class of interior solutions is obtained, with the help
of the NJA algorithm, with oblate spheroidal boundary surfaces.

\section{NJA Algorithm and the Choice of Coordinates}     
\indent The starting point of the NJA algorithm
is the static spherically symmetrical
seed metric . In spheroidal coordinates $t,r,\theta,\phi$
the seed metric is
\begin{equation}
ds^2=e^{2\mu(r)}dr^2+r^2(d{\theta}^2+{\sin}^2\theta d{\phi}^2)-e^{2\nu(r)}
dt^2.
\label{a1}
\end{equation} 
This metric can be expressed in terms of a null tetrad
$l_{\alpha},n_{\alpha},m_{\alpha},{\overline{m}}_{\alpha}$.
To make it short (for more details see \cite{7,11}), the NJA method
consists in taking a complex transformation of the metric (\ref{a1}), once
an appropriate coordinate system has been choosen. If Eddington-Finkelstein
coordinates are used with $dt=du-e^{\mu-\nu}dr$, these 
transformations become
\begin{equation}
u^{\prime}=u-\imath a\cos\theta\;,\;r^{\prime}=r+\imath a\cos\theta\;,\;
{\theta}^{\prime}=\theta\;,\;{\phi}^{\prime}=\phi .
\label{a2}
\end{equation}
In this way, the initial null tetrad relative to (\ref{a1}) acquires
complex components dependent on the constant $a$
(rotational parameter). In \cite{11} the authors
have been able to cast the metric in a simple form, with only a component
$g_{t\phi}$ out of diagonal, compatible with the Kerr metric
written in Boyer-Lindquist (BL) coordinates \cite{14}. 
Finally, in BL coordinates, the metric (\ref{a1}) becomes,
after the application of the NJA, 
\begin{eqnarray}
& &g_{tt}=-e^{2\nu(r,\theta)}\;,\;
g_{rr}=\frac{\Sigma}{\Sigma e^{-2\mu(r,\theta)}+a^2{\sin}^2\theta}\;,\;
g_{\theta\theta}=\Sigma\;,\nonumber\\
& &g_{\phi\phi}={\sin}^2\theta[\Sigma+a^2{\sin}^2\theta
e^{\nu(r,\theta)}(2e^{\mu(r,\theta)}-e^{\nu(r,\theta)})]\;,\nonumber\\
& &g_{t\phi}=a e^{\nu(r,\theta)}{\sin}^2\theta(e^{\mu(r,\theta)}
-e^{\nu(r,\theta)}),\nonumber\\
\label{a3}
\end{eqnarray}
where $\Sigma=r^2+a^2{\cos}^2\theta$ and $\nu,\mu$ now depend on $r,\theta$.
Setting 
\begin{equation}
e^{2\mu}=e^{-2\nu}=\frac{r^2+a^2{\cos}^2\theta}
{r^2-2rm+a^2{\cos}^2\theta},
\label{a4}
\end{equation}
we obtain the Kerr solution.
This can be obtained, in Cartesian
coordinates $(t,x,y,z)$, in the Kerr-Shild form:
\begin{eqnarray}
& &ds^2=dx^2+dy^2+dz^2-dt^2\nonumber\\
& &+\frac{2m}{r^4+a^2z^2}
{\left[\frac{r(xdx+ydy)+a(xdy-ydx)}{r^2+a^2}+\frac{z}{r}dz
+dt\right]}^2,\label{a6}
\end{eqnarray}
where the variable $r$  is defined as:
\begin{equation}
\frac{x^2+y^2}{r^2+a^2}+\frac{z^2}{r^2}=1.
\label{a7}
\end{equation}
Performing the coordinate transformation given by \cite{14}
\begin{eqnarray}
& &x=r\sin\theta\cos\phi+a \sin\theta\sin\phi, \nonumber\\
& &y=r\sin\theta\sin\phi-a \sin\theta\cos\phi, \nonumber\\
& &z=r\cos\theta\:\;,\;\;t\rightarrow t-2m\int\frac{rdr}{r^2-2mr+a^2},
\label{a8}
\end{eqnarray}
with the transformations (\ref{a8}), the metric takes exactly expression 
(\ref{a3}) with (\ref{a4}). From (\ref{a7}) we deduce that surfaces with
$r=const.$ are oblate spheroids that in the limit $a=0$ reduce to spheres.

\section{Matching Conditions and Generation of Kerr Interior Solutions
from Static Ones}
\indent Given two space-times with Lorentzian signature $(-+++)$, $M^{+}$
and $M^{-}$, with related metrics $g^{+}_{\alpha\beta}$ and
$g^{-}_{\alpha\beta}$ (for the space-time (\ref{a3}) we can use the same 
coordinate system both for $M^{+}$ and $M^{-}$), 
we can construct a new manifold $M$ by
joining the two space-times on a non-null 3-surface $S$,
with the help of the junction formalism
\cite{9,10}. To join $M^{+}$
and $M^{-}$ on $S$, we must calculate the metric tensor $g^{\pm}_{ij}$
on $S$ (first fundamental form) and the extrinsic curvature
(second fundamental form) $K_{ij}$. In terms of the unit normal to the 
surface $n^{\alpha}$ this is given by $K_{ij}=n_{i;j}$, where ``;'' denotes 
the covariant derivative and the equation for $S$ for non-null surfaces is
$f(x^{\alpha})=0$. Thus $n_{\gamma}=\pm \frac{{\partial}_{\gamma}f}
{\sqrt{{\partial}_{\beta}f{\partial}^{\beta}f}}$, the sign $\pm$
depending on the orientation of the 4-vector field $n_{\gamma}$.\\  
The DI conditions for $M^{+}$ and $M^{-}$ on $S$ are:
\begin{eqnarray}
& &[g_{ij}]=g^{+}_{ij}-g^{-}_{ij}=0\label{b1}\\
& &[K_{ij}]=K^{+}_{ij}-K^{-}_{ij}=0\label{b2}.
\end{eqnarray}
In what follows the subscript ``-''denotes the interior solution.\\
If both (\ref{b1}) and (\ref{b2}) are satisfied, then $S$ is a 
boundary surface. If only (\ref{b1}) is satisfied, $S$ is a thin shell
surface.
If we choose coordinates (\ref{a8}), on an oblate
spheroid with $r=R$ ($R$ a constant) 
for the line element (\ref{a3}), then the conditions (\ref{b1}) become
\begin{eqnarray}
& &{\left(e^{2{\nu}_{+}}\right)}_{S}=
   {\left(e^{2{\nu}_{-}}\right)}_{S}\;,\nonumber\\
& &{\left(e^{2{\mu}_{+}}\right)}_{S}=
   {\left(e^{2{\mu}_{-}}\right)}_{S}\;,\label{b3}
\end{eqnarray}
while conditions (\ref{b2}) are
\begin{eqnarray}
& &{\left({e^{2{\nu}_{+}}}_{,r}\right)}_{S}=
   {\left({e^{2{\nu}_{-}}}_{,r}\right)}_{S}\;,\nonumber\\
& &{\left({e^{2{\mu}_{+}}}_{,r}\right)}_{S}=
   {\left({e^{2{\mu}_{-}}}_{,r}\right)}_{S}\;,\label{b4}
\end{eqnarray}
where subindices ``,'' denote the partial derivative.\\
In terms of the Kerr exterior solution (\ref{a4}),
conditions (\ref{b3}) now look:
\begin{eqnarray}
& &{\left(e^{2{\nu}_{-}}\right)}_{S}=\frac{R^2+a^2{\cos}^2\theta-2mR}
{(R^2+a^2{\cos}^2\theta)}, \label{b5}\\
& &{\left(e^{2{\mu}_{-}}\right)}_S=\frac{R^2+a^2{\cos}^2\theta}
{(R^2+a^2{\cos}^2\theta-2mR)},\label{b6}
\end{eqnarray}
and (\ref{b4})
\begin{eqnarray}
& &{\left({e^{2{\nu}_{-}}}_{,r}\right)}_{S}=\frac{2m(R^2-a^2{\cos}^2\theta)}
{{(R^2+a^2{\cos}^2\theta)}^2},\label{b7}\\
& &{\left({e^{2{\mu}_{-}}}_{,r}\right)}_{S}=\frac{2m(a^2{\cos}^2\theta-R^2)}
{{(R^2+a^2{\cos}^2\theta-2mR)}^2}.\label{b8}
\end{eqnarray}
Our strategy for finding Kerr interior solutions is the following:
instead of starting with a stationary trial solution with conditions (\ref{b1})
(\ref{b2}), we start with a static physically
reasonable interior Schwarzschild solution. In this case we only need to 
know the explicit form of the static line element, but, in order to
find the pressure and the density profile of the star, it is not
necessary that the Volkov-Oppenheimer equation be solved analitically.
Further, it must be ensured that, starting with a trial stationary
solution with conditions (\ref{b1}) and (\ref{b2}), the static solution
is free of conical singularity: this imposes some additional conditions on
the metric components.
If a static seed  metric is chosen, the next step is to decide if the
surface $S$ is a boundary surface or a thin shell surface.
Finally, if the static seed metric is composed of elementary functions, we
can find the expression (depending on the parameter ``$a$'')
 that verifies
the appropriate junction conditions. 
In the next section we apply these reasonings to the imcompressible
perfect fluid Schwarzschild solution.

\section{Example One: The Incompressible Perfect Fluid 
Schwarzschild Solution}
\indent In \cite{6}, the authors first applied the NJA using
as seed metric the incompressible Schwarzschild solution.
However, they did not use the transformed metric in the simple form
(\ref{a3}), but rather applied the NJA algorithm performing a
complex transformation on the 
Schwarzschild interior solution
after taking the two constants of this solution as functions of $r$.
Consequently, the result they found in the
static limit represents an anisotropic fluid solution, not the starting
perfect fluid one.
Furthermore, this solution cannot satisfy the
strong energy condition.\\
The incompressible Schwarzschild solution  
is given by (\ref{a1}) with
\begin{eqnarray} 
& &e^{2\mu}=\frac{R^3}{R^3-2mr^2}, \label{c1}\\
& &e^{2\nu}={\left[\frac{3}{2}\sqrt{1-\frac{2m}{R}}
-\frac{1}{2}\sqrt{1-\frac{2mr^2}{R^3}}\right]}^2.\label{c2}
\end{eqnarray}
We have then (in appropriate units) 
$G_{\mu\nu}=T_{\mu\nu}$ and therefore for
the mass-energy density $\epsilon$ and for the radial pressure $p$ 
we obtain
\begin{eqnarray}
& &\epsilon=\frac{6m}{R^3},\label{c3}\\
& &p=\frac{\epsilon\left(\sqrt{1-\frac{2mr^2}{R^3}}-
\sqrt{1-\frac{2m}{R}}\right)}
{\left(3\sqrt{1-\frac{2m}{R}}-\sqrt{1-\frac{2mr^2}{R^3}}\right)}.
\label{c4}
\end{eqnarray}
Solutions (\ref{c1}) and (\ref{c2}) are stable under perturbations when
(Bondi limit \cite{Bondi}) $R>\frac{9}{4}m$.
The Schwarzschild interior solution is acausal, and thus the dominant energy
condition \cite{15} cannot be satisfied for the energy-momentum tensor
$T_{\mu\nu}$ of metric (\ref{a3}). Despite this, the unphysical
character becomes less important if this solution is considered as a
limiting case of the class of solutions with non-increasing
outwards density \cite{Buc}. Furthermore, 
the incompressible Schwarzschild solution
is the unique conformally flat, static perfect fluid solution and it
can be a model to describe the core of a neutron star.\\
First of all we must decide what boundary conditions must be imposed to
the stationary metric (\ref{a3}) with the seed metric (\ref{c1})
and (\ref{c2}), i.e. if only (\ref{b1}) or both (\ref{b1}) and (\ref{b2})
must be considered.\\
For (\ref{c1}), (\ref{c2}), the continuity of ${\nu}_{,r}(r)$ assures 
the continuity
of the hydrostatic pressure $p$ (\ref{c4}).
On the contrary , 
the continuity of ${\mu}_{,r}$ is not necessary because of the
equation:
\begin{equation}
\epsilon r^2=1-e^{-2{\mu}(r)}[1-2r{\mu}_{,r}].
\label{c6}
\end{equation}
In other words, the energy density is constant and thus cannot be 
vanishing on $r=R$.
In fact:
\begin{equation}
{\left({e^{2{\mu}_{+}}}_{,r}\right)}_S=\frac{-2m}{{(R-2m)}^2}
\neq \frac{4m}{{(R-2m)}^2}=
{\left({e^{2{\mu}_{-}}}_{,r}\right)}_S.
\label{c7}
\end{equation}
However, it must be pointed out that conditions (\ref{b7})
and (\ref{b8}) have been
obtained (see \cite{11}) assuming $a\neq 0$. When we derive condition
(\ref{b7})-(\ref{b8}), the parameter $a$ appears as an
overall factor in (\ref{b8}), and thus in the limit $a=0$ the only surviving
condition for the continuity of $K_{ij}$ is (\ref{b7}).\\
Conversely, in the stationary case with the line element (\ref{a3}), for the
continuity of $K_{ij}$ (boundary surface), both (\ref{b7}) and
(\ref{b8}) must be satisfied. Furthermore, 
if we consider the static limit as an 
analytic limit of stationary space-times, expression (\ref{c7})  says
that condition (\ref{b8}) cannot be satisfied when the parameter
$a$ is present. This fact can be confirmed 
performing the limit $a=0$ in (\ref{b8}), obtaining exactly expression
(\ref{c7}). Consequently, for the interior stationary
solutions generated by (\ref{c1}) and (\ref{c2}), only conditions
(\ref{b5}) and (\ref{b6}) must be imposed, for a thin shell
surface.\\ Moreover, it is possible to impose
to the stationary metric (\ref{a3}) the condition
(\ref{b7}) (satisfied for the chosen static seed solution ), but it appears
to be not necessary.
By inspection of (\ref{c1}), it is easy to see that the most general
expression for $e^{2\mu(r,\theta)}$ satisfying condition 
(\ref{b6}) is:                    
 \begin{equation}
e^{2{\mu}_{-}}=\frac{R^3+Ra^2{\cos}^2\theta}{R^3+Ra^2{\cos}^2\theta-2mr^2}
+B(r,\theta,a^2),
\label{c8}
\end{equation}	
where $B(r,\theta,a^2)$ is an arbitrary function such that:
\begin{equation}
B(R,\theta,a^2)=B(r,\theta,0)=0.
\label{c9}
\end{equation}
Obviously, because of the axial simmetry of (\ref{a3}), only even powers in
$a$ are allowed, and $B(r,\theta,a^2)$ must be such that expression
(\ref{c8}) is non-negative.
If $B(r,\theta,a^2)$ is a regular function of its arguments, then expression
(\ref{c8}) is regular for $R>2m$. 
Finally, the most general
expression for $e^{2\nu(r,\theta)}$ that satisfies condition (\ref{b5}) is:
\begin{eqnarray}
& &e^{2{\nu}_{-}}=\frac{5}{2}-\frac{9mR}{2(R^2+a^2{\cos}^2\theta)}-
\frac{mr^2}{2R(R^2+a^2{\cos}^2\theta)}\nonumber\\
& &-\frac{3}{2}\sqrt{1-\frac{2mR}{(R^2+a^2{\cos}^2\theta)}}
\sqrt{1-\frac{2mR(r^2+a^2{\cos}^2\theta)}
{{(R^2+a^2{\cos}^2\theta)}^2}}+H(r,\theta,a^2),\label{c10}
\end{eqnarray}
where $H(r,\theta, a^2)$ is an arbitrary regular function with
\begin{equation}
H(R,\theta,a^2)=H(r,\theta,0)=0 
\label{ce}
\end{equation}
and such that
expression (\ref{c10}) is non-negative. Even in this case, 
(\ref{c10}) is regular for $R>2m$. If we impose condition
(\ref{b7}), we have:
\begin{eqnarray}
& &e^{2{\nu}_{-}}= H_{\alpha}(a^2,r,\theta)+ 
\left[ \frac{5}{2}-\frac{9mR}{2(R^2+a^2{\cos}^2\theta)}-
\frac{mr^2}{2R(R^2+a^2{\cos}^2\theta)}\right. \nonumber\\ 
& &\left.-\frac{3}{2}\sqrt{1-\frac{2mR}{(R^2+a^2{\cos}^2\theta)}}
\sqrt{1-\frac{2mR(r^2+a^2{\cos}^2\theta)}
{{(R^2+a^2{\cos}^2\theta)}^2}}\right]e^{\alpha(a^2,r,\theta)},\label{c100}
\end{eqnarray}
where $\alpha$ is a regular function such that 
$\alpha(0,r,\theta)=\alpha(a^2,R,\theta)=0$. Also
$H_{\alpha}$ is a regular vanishing function for $a=0$ and  
$r=R$ and depends on the choice made for $\alpha$. For example, setting
$\alpha =0$, we have:
\begin{equation}
H_{0}(r,\theta,a^2)=\frac{c_H H(r)a^2{\cos}^2\theta}
{{(R^2+a^2{\cos}^2\theta)}^2}, \label{c11}
\end{equation}
where $H(r)$ is a regular function such that $H(R)=0$ and the constant
$c_H$ in (\ref{c11}) depends on $H(r)$. For example, if
$H(r)=(R^2-r^2)$ then $c_H=\frac{m}{2R}$.\\
Expressions (\ref{c8}) and (\ref{c10}), when substituted in (\ref{a3})
give our class of stationary interior solutions describing
an anisotropic fluid.\\ 
In the literature, the only 
known exact anisotropic fluid solution is given in \cite{23}.
In the next section we study the solutions so obtained at the first order
in the rotational parameter $a$. 

\section{Slowly Rotating Limit}
\indent The suitable metric  to describe a slowly rotating star is given by
\cite{24}
\begin{equation}
ds^2=e^{2\mu(r,\theta)}dr^2+
e^{2{\mu}_2(r,\theta)}d{\theta}^2+
e^{2\psi(r,\theta)}{[d\phi+\omega(r,\theta)dt]}^2
-e^{2\nu(r,\theta)}dt^2
\label{d1}
\end{equation}
where, with $\Omega=\frac{d\phi}{dt}$ 
denoting the angular velocity as seen from a rest observer at
spatial infinity, we have:
\begin{eqnarray}
& &\nu={\nu}_0+o({\Omega})\;\;\;,\;\;\;\psi={\psi}_0+
o({\Omega})\nonumber\\
& &\mu={\mu}_0+o({\Omega})\;\;\;,\;\;\;{\mu}_2={\mu}_{20}+
o({\Omega}).\label{d2}
\end{eqnarray}
Here ${\nu}_0,{\psi}_{0}\cdots$ represent the static metric functions.
By slow rotation we mean that the star rotates with an angular velocity
$\Omega$ so small that distorsions to its shape of order
${\Omega}^2$ can be ignored.
The metric (\ref{d1}), at order $\Omega$, reduces to:
\begin{equation}
ds^2=e^{2\mu(r)}dr^2+r^2(d{\theta}^2+{\sin}^2\theta d{\phi}^2)+
2\omega(r){\sin}^2\theta dt d\phi-e^{2\nu(r)}dt^2,
\label{d3}
\end{equation}
with $\mu$ and ${\nu}$ given by (\ref{c1}) and (\ref{c2}) and
$\omega(r)=a e^{\nu}(e^{\mu}-e^{\nu})$.\\
Therefore, at the boundary $R$ we have 
$\omega(R)=\frac{2m}{R}$, as expected for a slowly rotating
star in the vacuum.\\
Note that since at order $\Omega$ the star is spherical
(the distorsion to its spherical shape appears at 
order ${\Omega}^2$, 
where the centrifugal force comes in action), we have
$\Omega=\frac{5a}{2R^2}$(the inertial momentum $I$ of a
sphere is $I=\frac{2}{5}mR^2$).
With the metric (\ref{d3}), the only first
order component of the Einstein's tensor $G_{\mu\nu}$ is
$G_{t\phi}$. The other non-vanishing components  are zero order
expressions in $a$ and  are thus the same of the static metric.
Then, the energy-momentum tensor 
at the order $a$ has the $tt,rr,\theta\theta,\phi\phi$
components
of the metric (\ref{c1}) and (\ref{c2}) of a perfect fluid with
\begin{eqnarray}
& &T_{\mu\nu}=(\epsilon+p)u_{\mu}u_{\nu}+pg_{\mu\nu},\nonumber\\
& &u^t=\frac{dt}{ds}=e^{-\nu}\;\;,\;\;u^{\theta}=u^r=0\;\;,\;\;
u^{\phi}=\frac{d\phi}{ds}=\frac{5a}{2R^2}e^{-\nu}.\label{d4}
\end{eqnarray}
Conversely, the equation involving $G_{t\phi}$ is
\begin{eqnarray}
G_{t\phi}&=&{\sin}^2\theta e^{-2\mu}[\frac{\omega}{r^2}+
\frac{{\omega}_{,r}{\mu}_{,r}}{2}+\frac{{\omega}_{,r}{\nu}_{,r}}{2}-
\frac{2\omega{\mu}_{,r}}{r}+ \nonumber\\
&+&\omega{\nu}^{2}_{,r}-\omega{\nu}_{,r}{\mu}_{,r}-
\frac{{\omega}_{,r,r}}{2}+\omega{\nu}_{,r,r}]+o(a),\label{d5}
\end{eqnarray}
Thus, once $\nu$ and $\mu$ are substituted with (\ref{c1}) and (\ref{c2}),
we get:
\begin{equation}
G_{t\phi}=T_{t\phi}={\sin}^2\theta\left[-r^2
\epsilon\left(\frac{\omega}{r^2}+\frac{5a}{2R^2}\right)-
\frac{5ap r^2}{2R^2}+{\Phi}_{t\phi}(r)\right]+o(a).
\label{d6}
\end{equation}
If the star, at order $a$, was described by a perfect fluid, then
the quantity ${\Phi}_{t\phi}$ would be identically 0.
This non-zero component makes the fluid  a non-perfect one.
The energy-momentum tensor in the slowly
rotating approximation has the same diagonal components of the static
Schwharzschild one  with a component $(t\phi)$ out of diagonal composed
of a perfect fluid part and anisotropic part given
by ${\Phi}_{t\phi}$: the energy
density $\epsilon$ and the radial pressure $p$ are again given by
(\ref{c3}) and (\ref{c4}).\\
Concerning the energy conditions, for a tensor of the type considered
(type I, see \cite{15}), denoting with ${\lambda}_0$ the eigenvalue
corresponding to the timelike eigenvector, and with ${\lambda}_i$
($i=1,2,3$) the eigenvalues corresponding to spacelike
eigenvectors, we have for the weak energy condition:
\begin{equation}
-{\lambda}_0\geq 0\;\;,\;\;-{\lambda}_0+{\lambda}_i\geq 0;
\label{d7}
\end{equation}   
for the strong energy condition:
\begin{equation}
-{\lambda}_0+{\sum}{\lambda}_i \geq 0\;\;\;,\;\;\;
-{\lambda}_0+{\lambda}_i\geq 0;
\label{d8}
\end{equation}
and for the dominant energy condition:
\begin{equation}
-{\lambda}_0\geq 0\;\;\;,\;\;\;{\lambda}_0\leq {\lambda}_i
\leq -{\lambda}_0,
\label{d9}
\end{equation}
where the eigenvalues are solutions of the equation:
\begin{equation}
|T_{\alpha\beta}-\lambda g_{\alpha\beta}|=0.
\label{d10}
\end{equation}
It is a simple matter to verify that at order $a$ the equation
(\ref{d10}) is the same as the static limit, and thus (\ref{d7}) and
(\ref{d8}) are satisfied (remember that the seed static solution
considered is acausal). In particular we have:
\begin{equation}
{\lambda}_0=-\epsilon +o(a)\;\;\;,\;\;\;{\lambda}_i=p(r)+o(a).
\label{d11}
\end{equation}
In the next section we collect some remarks on the energy conditions for
our solution .

\section{Considerations on Energy Conditions}
\indent The eigenvalue equation (\ref{d10}) for the metric 
(\ref{a3}), becomes:\\
\\$\begin{vmatrix}
T_{tt}-\lambda g_{tt}&0&0&T_{t\phi}-\lambda g_{t\phi}\\
0&T_{rr}-\lambda g_{rr}&T_{r\theta}&0\\
0&T_{r\theta}&T_{\theta\theta}-\lambda g_{\theta\theta}&0\\
T_{t\phi}-\lambda g_{t\phi}&0&0&T_{\phi\phi}-\lambda g_{\phi\phi}
\end{vmatrix}\;=\;0$\\
\\and the eigenvalues are thus given by:
\begin{eqnarray}
& &{\lambda}_{0,1}=\frac{P\pm\sqrt{P^2-4(g_{tt}g_{\phi\phi}-
g_{t\phi}^2)(T_{tt}T_{\phi\phi}-T_{t\phi}^2)}}
{2[g_{tt}g_{\phi\phi}-g_{t\phi}^2]}, \nonumber\\
& &P=(T_{tt}g_{\phi\phi}+g_{tt}T_{\phi\phi}-2T_{t\phi}g_{t\phi}),\label{e1}\\
& &{\lambda}_{2,3}=\frac{Q\pm\sqrt{Q^2-4g_{rr}g_{\theta\theta}
(T_{rr}T_{\theta\theta}-T_{r\theta}^2)}}
{(2g_{rr}g_{\theta\theta})},\nonumber\\
& &Q=(T_{rr}g_{\theta\theta}+g_{rr}T_{\theta\theta}).\label{e2}
\end{eqnarray}
First of all, note that it is not possible to perform a perturbative
expansion with respect to the parameter $a$. In fact, the metric 
(\ref{a3}) with (\ref{c8}) and (\ref{c10}) is not continuous in the 
variables $(r,a)$ in all the interior shell. For example, if we take
the metric element $g_{rr}$ with (\ref{c8}) and (\ref{c9}), then the limits
for $a\rightarrow 0, r\rightarrow 0$ cannot be exchanged. Further,
if we expand formally the component $g_{rr}$ up to the order $a^2$, we get:
\begin{eqnarray}
& &g_{rr}=\frac{R^3}{R^3-2mr^2}+\nonumber\\
& &\frac{a^2[-2mr^4 R{\cos}^2\theta+
B(r^2 R^6-4r^4 R^3 m+4r^6 m^2)-R^6{\sin}^2\theta]}
{r^2{(R^3-2mr^2)}^2}.\label{e3}
\end{eqnarray}
If we take $B(r,\theta)$ to be a regular function at $r=0$, then expression 
(\ref{e3}) is divergent to $-\infty$, whereas the component $g_{rr}$,
given by (\ref{a3}) with (\ref{c8}),
is bounded in a neighbourhood of $r=0$. Therefore, a perturbative expansion
in the parameter ``$a$'' is possible only in the slowly rotating limit,
where the functions $\mu(r, \theta)$ and $\nu(r, \theta)$ are the static
ones.\\
The study of the behaviour of the
metric in $r=0$ is important for the energy conditions since a singularity
can arise at such a point.
Expressions (\ref{e1}) and (\ref{e2}) show
a complicated dependence on $H(r,\theta,a^2), B(r,\theta,a^2)$.
Therefore, the study of (\ref{d7})-(\ref{d9}) is not a simple task.
However, some remarks can be made. First of all, the components
$T_{tt}, T_{t\phi}, T_{rr}, T_{r\theta}, T_{\theta\theta}$ have a finite limit
on the axis at $\theta=0,\pi$, and for $T_{\phi\phi}$ we have
$T_{\phi\phi}\sim {\sin}^2\theta$. Therefore the eigenvalues 
(\ref{e1}) and (\ref{e2}) are well defined on the axis where the norm
of the spacelike Killing vector ${\xi}_{\phi}$ is vanishing. Further,
the metric functions are regular when $R>2m$. The only singularity can arise 
when the determinant of the metric is vanishing, i.e. at 
$r=0, \theta=\frac{\pi}{2}$ (ring singularity).\\ 
Concerning the energy conditions, the inequality 
${-\lambda}_0\geq 0$ is satisfied
if (for example) 
$T_{tt}T_{\phi\phi}-T_{t\phi}^2\geq 0$ (with $g_{\phi\phi}\geq 0,
g_{tt}\leq 0$) and  
consequently $-{\lambda}_0+{\lambda}_1\geq 0$ is satisfied.
The energy conditions cannot be satisfied if ${\lambda}_2, {\lambda}_3$ have
a large divergent negative part and ${\lambda}_0$ with a less negative
 divergent part. The stress-energy tensor has only ($R>2m$) a ring 
singularity at $r=0, \theta=\frac{\pi}{2}$. In
a neighbourhood of the ring, the most rapidly diverging part 
looks as follows (with
$\Sigma=r^2+a^2{\cos}^2\theta$):
\begin{eqnarray}
& &T_{tt}\sim {\Sigma}^{-4}\;,\;T_{t\phi}\sim {\Sigma}^{-4}\;,\;
T_{\phi\phi}\sim {\Sigma}^{-4},\nonumber\\
& &T_{rr}\sim {\Sigma}^{-2}\;,\;T_{\theta\theta}\sim {\Sigma}^{-2}\;,\;
T_{r\theta}\sim {\Sigma}^{-1}.\label{e4}
\end{eqnarray}   
Therefore, we can find the general expressions needed for
$B(r,\theta,a^2), H(r,\theta,a^2)$ such that we can reduce the most
rapidly divergent part in (\ref{e1}) and (\ref{e2}). We can take:
\begin{eqnarray}
& &B(r,\theta,a^2)={\Sigma}^k[A(r,\theta,a^2){(R-r)}^
{\epsilon}+L(r,\theta,a^2)(R^{\gamma}-r^{\gamma})]+\nonumber\\
& &C(r,\theta,a^2){(R-r)}^{\delta}+
D(r,\theta,a^2)(R^{\Delta}-
r^{\Delta})+\nonumber\\
& &{[{\Sigma}^{\psi}-{(R^2+a^2{\cos}^2\theta)}^{\psi}]}^{\beta}
E(r,\theta,a^2),\label{e5}
\end{eqnarray}
where $(\epsilon, \gamma, \delta, \Delta, \psi, \beta)$ are non-negative
constants (assuming the regularity of the
metric coefficients) and the functions 
$A, L, C, D, E$ are regular functions of
their arguments and vanish for $a=0$. A similar expression follows for
$H(r,\theta, a^2)$ with obviously different constants
$({\epsilon}^{\prime}, {\gamma}^{\prime}\cdots)$ and functions
$A^{\prime}, L^{\prime}\cdots$. The next step is to choose the 
parameters and functions in (\ref{e5}) in such a way that 
$-{\lambda}_0\geq 0$ and that ${\lambda}_0$ has a divergent part at most of 
order $\sim {\Sigma}^{-1}$: this way the value of the volume integral of 
${\lambda}_{0}$ on $r\leq R$ is certainly bounded. Furthermore,
the eigenvalues (\ref{e2}) can be divergent on the ring, but with 
a positive divergent part. Moreover, the acausality of the chosen
seed metric  does not imply that conditions (\ref{d9}) cannot be satisfied.
Finally, the hydrostatic pressure must be vanishing at the boundary
$r=R$.\\
As stated above, expressions (\ref{e1}) and (\ref{e2}) are too long
and complicated to easily find the correct expressions for
$B(r,\theta,a^2)$ and $H(r,\theta,a^2)$ (if they exist!).\\

\section{Example Two: Conformally Flat Anisotropic Seed Metric}
\indent In this last section we give an example of a 
class of Kerr interior solutions
with a more appealing geometry than the one considered
above, i.e. with boundary surfaces and therefore without distributional
source at $r=R$.
To this purpose, we start with the class of static anisotropic
conformally flat
solutions found by Stewart (see \cite{st}). The starting point is the 
energy-momentum tensor $T_{\mu\nu}$ in the form (see \cite{let}):
\begin{equation}
T_{\mu\nu}=(\epsilon+p)u_{\mu}u_{\nu}+pg_{\mu\nu}+
(\sigma-p){\xi}_{\mu}{\xi}_{\nu},
\label{f1}
\end{equation}
where $u_{\mu}$ is the anisotropic fluid four-velocity, ${\xi}_{\mu}$ is
a spacelike unit four-vector pointing in the direction of the
anisotropy, $\epsilon$ is the rest energy density, $\sigma$ is the 
pressure along the anisotropy and $p$ is the pressure
on a plane perpendicular to the anisotropy direction.\\
A class of static solutions, matching at the boundary with the vacuum
Schwarzschild metric, can be obtained with:
\begin{equation}
\epsilon = \frac{6m}{R^3(1-\frac{3}{5}q)}\left(1-\frac{qr^2}{R^2}\right)\;\;\;,\;\;\;0\leq q \leq 1 .
\label{f2}
\end{equation}
For $q=0$ we ragain the incompressible Schwarzschild solution.
As we have seen in section 2, to obtain solutions with
boundary surfaces, we need a $C^1$ expression for $e^{2\mu(r)}$.
This can be obtained if and only if we set $q=1$ ($\epsilon(R)=0$).
Consequently, the solution ($q=1$), with (\ref{a1}), is:
\begin{eqnarray}
& &e^{2\nu(r)}=e^{2\Phi(r)}{\left(cr^2 e^{-2\Phi(r)}+b\right)}^2 ,\label{f3}\\
& &e^{2\mu(r)}=\frac{R^3}{(R^3-5mr^2+\frac{3mr^4}{R^2})} ,\nonumber\\
& &e^{2\Phi(r)}=\frac{1}{2}\left[1+\sqrt{1-\frac{5mr^2}{R^3}+
\frac{3mr^4}{R^5}}-\frac{5mr^2}{2R^3}\right] ,\nonumber\\
& &c=\frac{1}{2}\frac{e^{\Phi(R)}}{R^2}\left[\frac{3m}{R}-1+
\sqrt{1-\frac{2m}{R}}\right] ,\nonumber\\
& &b=\frac{1}{2}e^{-\Phi(R)}\left[1-\frac{3m}{R}+
\sqrt{1-\frac{2m}{R}}\right].\nonumber
\end{eqnarray}  
Note that both the metric
metric functions $g_{tt}$ and $g_{rr}$ are $C^1$ on the 
boundary at $r=R$. Therefore, with the help of the procedure discussed above,
we can write a class of interior Kerr solutions satisfying the conditions
(\ref{b5})-(\ref{b8}), i.e. the continuity of the first and the second
fundamental form (boundary surfaces). The solutions so obtained are: 
\begin{eqnarray}
& &e^{2{\mu}_{-}}=\frac{[R^3+Ra^2{\cos}^2\theta]e^{A(a^2,r,\theta)}}
    {(R^3+Ra^2{\cos}^2\theta-5mr^2+\frac{3mr^4}{R^2})}+
     S_A(a^2,r,\theta),\label{f8}\\
& &e^{2\Phi(a^2,r,\theta)}=\frac{1}{2}\left[1-\frac{5mr^2}
      {2R(R^2+a^2{\cos}^2\theta)}\right.\nonumber\\
& &\left.+\sqrt{1-\frac{5mr^2}{R(R^2+a^2{\cos}^2\theta)}+
\frac{3mr^4}{R^3(R^2+a^2{\cos}^2\theta)}}\right],\nonumber\\
& &e^{2{\nu}_{-}}=e^{2\Phi(a^2,r,\theta)}
{\left[cr^2e^{-2\Phi(a^2,r,\theta)}+b\right]}^2
 e^{B(a^2,r,\theta)}+H_B(a^2,r,\theta),\nonumber\\
& &c=\frac{1}{2R^2}e^{\Phi(a^2,R,\theta)}
\left[\frac{3mR}{R^2+a^2{\cos}^2\theta}
-1+\sqrt{1-\frac{2mR}{R^2+a^2{\cos}^2\theta}}\right], \nonumber\\
& &b=\frac{1}{2}e^{-\Phi(a^2,R,\theta)}
\left[1-\frac{3mR}{R^2+a^2{\cos}^2\theta}+ 
\sqrt{1-\frac{2mR}{R^2+a^2{\cos}^2\theta}}\right],\nonumber
\end{eqnarray}
where $A(a^2,r,\theta)$ and $B(a^2,r,\theta)$ are arbitrary regular functions
such that:
\begin{equation}
A(0,r,\theta)=A(a^2,R,\theta)=B(0,r,\theta)=B(a^2,R,\theta)=0
\label{f13}
\end{equation}
and (see discussion before equation (\ref{c100})) $S_A$ and
$H_B$ are regular vanishing functions for $a=0$ and $r=R$
and depend on the choice made for $A$ and $B$.\\
Setting regular expressions for $A$ and $B$, the metric so obtained
is regular for $R>\frac{25}{12}m$ and on the symmetry axis, with the
exception of the ring singularity on the equatorial plane for
$r=0, \theta=\frac{\pi}{2}$.\\
Also in this case, since the expressions involved are very long and
show a complicated dependence on $A, S_A, B, H_B$, 
the study of energy conditions is not a simple
task. However, the energy-momentum for (\ref{f8}) is again
of type I, and the reasonings made in section 5 are again valid.\\
As a final remark, we point out that the procedure discussed in this paper
can be applied not only for obtaining Kerr interior solutions but also
to find interior solutions matching with a general 
asymptotically flat vacuum stationary spacetime, provided that we start
with a known reasonable static metric.


\begin{thebibliography}{99}

\bibitem{1}R. P. Kerr, {\it Phys. Rev. Lett.} {\bf 11}, 237 (1963). 

\bibitem{2}P. Florides, {\it Nuovo Cimento} {\bf B 13}, 1 (1973).

\bibitem{3}P. Florides and J. Synge, {\it Proc. Roy. Soc.}
{\bf A 280}, 459 (1964).  

\bibitem{4}M. Gurses and F. Gursey, {\it J. Math Phys.}
{\bf 16}, 2385 (1975).

\bibitem{5}H. Wahlquist, {\it Phys. Rev.} {\bf 172, 5}, 1291 (1968).
 
\bibitem{6}L. Herrera and L. Jimenez, {\it J. Math Phys.} {\bf 23(12)}, 2339 (1982). 

\bibitem{7}E. T. Newman and A. Janis, {\it J. Math. Phys.} {\bf 6}, 915 (1965). 
\bibitem{8}R. H. Boyer and T. G. Price, {\it Proc. Camb. Phil. Soc.}
{\bf 61}, 531 (1965).

\bibitem{9}W. Israel, {\it Nuovo Cimento} {\bf 44}, 1 (1966).  

\bibitem{10}W. Israel, {\it Phys. Rev. D} {\bf 2}, 641 (1970).

\bibitem{11}S. P. Drake and R. Turolla, {\it Class. Quantum Grav.}
{\bf 14}, 1883 (1997). 

\bibitem{12}S. P. Drake and Szekeres, {\it Gen. Rel. Grav.} {\bf 32}, 445 (2000) ({\it Preprint} gr-qc/9807001). 

\bibitem{13}J. R. Oppenheimer and G. Volkov, {\it Phys. Rev.} {\bf 55}, 374 (1939).

\bibitem{st}B. W. Stewart, {\it J. Phys. A: Math Gen} {\bf 15}, 2419 (1982).

\bibitem{14}R. Boyer and R. W. Lindquist, {\it J. Math. Phys.} {\bf 8}, 265 (1967).

\bibitem{Bondi}H. Bondi, {\it Proc. R. Soc. A} {\bf 282}, 303 (1964).

\bibitem{15}S. W. Hawking and G. F. R. Ellis, {\it The Large Scale
strutcure of space-time} (Cambridge: Cambridge University Press, 1973).

\bibitem{Buc}H. A. Buchdahl, {\it Phys. Rev.} {\bf 116}, 1027 (1959).

\bibitem{23}T. Papakostas, {\it Int. J. Mod. Phys. D} {\bf 10}, 869 (2001).

\bibitem{24}J. B. Hartle, {\it Astrophys. J.} {\bf 150}, 1005 (1967).

\bibitem{let}P. S. Letelier, {\it Phys. Rev. D} {\bf 22}, 807 (1980).
 
\end{thebibliography}
\end{document}